\documentclass[epj]{svjour}

\usepackage{amssymb}
\usepackage{amsmath}
\usepackage{graphicx}
\usepackage{hyperref}
\usepackage{longtable}

\begin{document}

\title{Revisiting Generalized Chaplygin Gas as a Unified Dark Matter and Dark Energy Model}

\author{Lixin Xu\inst{1,2,3} \and Jianbo Lu \inst{4}\and Yuting Wang\inst{1,5}}


\institute{Institute of Theoretical Physics, Dalian University of Technology, Dalian,
116024, People's Republic of China \and College of Advanced Science \& Technology, 
Dalian University of Technology, Dalian, 116024, People's Republic of China \and Korea Astronomy and Space Science Institute,
Yuseong Daedeokdaero 776,
Daejeon 305-348,
R. Korea \and Department of Physics, Liaoning Normal University, Dalian 116029, People's Republic of China \and Institute of Cosmology \& Gravitation,
University of Portsmouth, Portsmouth, PO1 3FX, United Kingdom}

\abstract{
In this paper, we revisit generalized Chaplygin gas (GCG) model as a unified dark matter and dark energy model. The energy density of GCG model is given as $\rho_{GCG}/\rho_{GCG0}=[B_{s}+(1-B_{s})a^{-3(1+\alpha)}]^{1/(1+\alpha)}$, where $\alpha$ and $B_s$ are two model parameters which will be constrained by type Ia supernova as standard candles, baryon acoustic oscillation as standard rulers and  the seventh year full WMAP data points. In this paper, we will not separate GCG into dark matter and dark energy parts any more as adopted in the literatures. By using  Markov Chain Monte Carlo method, we find the result: $\alpha=0.00126_{-    0.00126-    0.00126}^{+    0.000970+    0.00268}$ and $B_s= 0.775_{-    0.0161-    0.0338}^{+    0.0161+    0.0307}$.
}



\maketitle

\section{Introduction}

 Since 1998, the type Ia supernova (SNIa) observations \cite{SNIa98}
 have shown that our universe has entered into a
phase of accelerating expansion. During these years from that time,
many additional observational results, including current Cosmic
Microwave Background (CMB) anisotropy measurement \cite{CMB03}, and
the data of the Large Scale Structure (LSS) \cite{LSS04}, also
strongly support this suggestion. And these cosmic observations
indicates that baryon matter component is about 4\% for total energy
density, and about 96\% energy density in universe is invisible. Considering the four-dimensional  standard cosmology, this accelerated expansion for universe predict that dark energy (DE) as
an exotic component with negative pressure is filled in universe. And it is shown that DE takes up about two-thirds of the total energy density from cosmic observations. The remaining one third is dark matter (DM). In theory mounting  DE models have already been constructed, for the reviews please see \cite{ref:DEReview1,ref:DEReview2,ref:DEReview3,ref:DEReview4,ref:DEReview5,ref:DEReview6,ref:DEReview7}. But there exists another possibility that the invisible energy component is a unified dark fluid. i.e. a mixture of dark matter and dark energy. Actually, based on the Einstein's gravity theory and the observed geometry structure of our universe, one can define the so-called dark fluid 
\begin{equation}
T^{dark}_{\mu\nu}=\frac{1}{8\pi G}G_{\mu\nu}-T^{obs}_{\mu\nu}
\end{equation}
where $G_{\mu\nu}$ is the Einstein tensor and $T^{obs}_{\mu\nu}$ is the observed energy-momentum tensor. This property is dubbed as dark degeneracy \cite{ref:darkdeneracy,ref:darkdeneracyxu}.

In these unified dark fluid models, the Chaplygin gas (CG) and its generalized model
have been widely studied for interpreting the accelerating universe
\cite{GCG,GCG-action,GCGpapers,GCGdecomp}. The most interesting property for this
scenario is that, two unknown dark sections in universe--dark energy
and dark matter  can be unified by using an exotic equation of
state. The original Chaplygin gas  model can be obtained from the
string Nambu-Goto action in the light cone coordinate
\cite{CG-string}. For generalized Chaplygin gas (GCG), it emerges as
an effective fluid of a generalized d-brane in a $(d+1, 1)$ space
time, and its action can be written as a generalized Born-Infeld
form \cite{GCG-action}.
  Considering that the application of string theory in principle is
in very high energy when the quantum effects is important in early
universe \cite{CG-string}, the quantum cosmological studies of the
CG and the GCG has been well investigated in Ref. \cite{CG-string}
and \cite{GCG-quantum}.

The GCG model is characterized by two model parameters $B_s$ and $\alpha$. To constrain the model parameter space, the GCG model has been confronted by cosmic observations, please see \cite{ref:GCGconstraint,ref:GCGGRBs,ref:GCGunified} for examples. In Ref. \cite{ref:GCGconstraint}, the geometric information from SN Ia, the baryon acoustic oscillation (BAO) and shift parameter $R$, $l_a(z_\ast)$ and $z_\ast$ from cosmic microwave background radiation (CMB) were used, where $B_s=0.73^{+0.06}_{-0.06}$ and $\alpha=-0.09^{+0.15}_{-0.12}$ was obtained. In these papers, the GCG was decomposed into two parts: effective dark matter and dark energy. And the effective dimensionless matter energy density was given as $\Omega_m=\Omega_b+(1-\Omega_b)(1-B_s)^{1/(1+\alpha)}$, where $B_s$ and $\alpha$ are dimensionless model parameters, for the details please see Eq. (\ref{eq:mcg}) of this paper. In Ref. \cite{ref:GCGGRBs}, we obtained the results: $B_s=0.7475^{+0.0556}_{-0.0539}$ and $\alpha=-0.0256^{+0.1760}_{-0.1326}$, where high redshift Gamma ray bursts data points were added as 'standard candles' and the decomposition was also employed. In these papers, the perturbation of GCG was not considered. So the values of model parameter $\alpha<0$ was included. In the following sections of this paper, we will see that the values of $\alpha\ge 0$ are mandatory for the stabilities of GCG perturbations. As is pointed by the authors in \cite{GCGdecomp}, when the phantom region is forbidden, one has a unique decomposition of GCG. However, the possibility of phantom dark energy has not been ruled out by current cosmic observations \cite{ref:wmap7}. Currently there is still a lack of physics principle to do any decomposition. In fact one can list many kinds of decompostions, for example $\rho_c=\rho_{c0}a^{-3}$ defined as cold dark matter and the remaining part $\rho_{de}=\rho_{GCG}-\rho_{c}$ defined as dark energy. The serious problem would be that there does not exist any decomposition at all, because it is a whole energy component. So any kind of decomposition would be non-proper. In Ref. \cite{ref:GCGunified}, the authors took GCG as a whole energy component, where the perturbations of this unified dark fluid was also discussed by WMAP-5 data and matter (baryon) power spectrum. In that paper, a tighter constraint was obtained when the matter (baryon) power spectrum was included. Based on these points, in this paper, we will use the currently available observational data sets: the full information of 7-year WMAP CMB data, BAO and the Union2 SNIa data to constrain the GCG model as the unification of dark matter and dark energy. We will see that when the full information of CMB is included, a tighter constraint will be obtained. 

This paper is organized as follows. In section II, the GCG model as the unification of dark matter and dark energy is introduced briefly. Based on the   observational data, we constrain the GCG model in section III. Section IV is the summary.

\section{Main equations in generalized Chaplygin gas model}   \label{sec:GCG}

The equation of state (EoS) of GCG model is given in the form of
\begin{equation}
p_{GCG}=-A/\rho^{\alpha}_{GCG}
\end{equation}
where $A$ and $\alpha$ are model parameters. In general, for a spatially non-flat FRW universe, the metric is
\begin{equation}
ds^{2}=-dt^{2}+a^{2}(t)\left[\frac{1}{1-kr^{2}}dr^{2}+r^{2}(d\theta^{2}+\sin^{2}\theta d\phi^{2})\right],
\end{equation}
where $k=0,\pm 1$ is the three-dimensional curvature and $a$ is the scale factor. Using the energy conservation of GCG, one can rewrite its energy density as
\begin{equation}
\rho_{GCG}=\rho_{GCG0}\left[B_{s}+(1-B_{s})a^{-3(1+\alpha)}\right]^{\frac{1}{1+\alpha}}\label{eq:mcg}
\end{equation}
where $B_{s}=A/\rho^{1+\alpha}_{GCG0}$ and
$\alpha$ are model parameters. Form Eq. (\ref{eq:mcg}),
one can find that $0\le B_s \le 1$ is demanded to keep the
positivity of energy density. If $\alpha=0$ in Eq. (\ref{eq:mcg}),
the standard $\Lambda$CDM model is recovered. Taking GCG as a
unified component, one has the Friedmann equation
\begin{eqnarray}
H^{2}&=&H^{2}_{0}\left\{(1-\Omega_{b}-\Omega_{r}-\Omega_{k})\left[B_{s}+(1-B_{s})a^{-3(1+\alpha)}\right]^{\frac{1}{1+\alpha}}\right.\nonumber\\
&+&\left.\Omega_{b}a^{-3}+\Omega_{r}a^{-4}+\Omega_{k}a^{-2}\right\}
\end{eqnarray}
where $H$ is the Hubble parameter with its current value $H_{0}=100h\text{km s}^{-1}\text{Mpc}^{-1}$, and $\Omega_{i}$ ($i=b,r,k$) are dimensionless energy parameters of baryon, radiation and effective curvature density respectively. In this paper, we only consider the spatially flat FRW universe.

To study the effects on CMB anisotropic power spectrum, the perturbation evolution equations for GCG would be  studied. We treat GCG as a unified dark fluid which interacts with the remaining matter purely through gravity. With assumption of pure adiabatic contribution to the perturbations, the speed of sound for GCG is
\begin{equation}
c^{2}_{s}=\frac{\delta p}{\delta \rho}=\frac{\dot p}{\dot \rho}=-\alpha w,\label{eq:cs2}
\end{equation}
where $w$ is the EoS of GCG in the form of
\begin{equation}
w=-\frac{B_{s}}{B_{s}+(1-B_{s})a^{-3(1+\alpha)}}.
\end{equation}
From the above equation, one can find that the values of $w$ are non-positive. So to keep the non-negativity of sound of speed, $\alpha\ge 0$ is required.

In the synchronous gauge, using the conservation of energy-momentum tensor $T^{\mu}_{\nu;\mu}=0$, one has the perturbation equations of density contrast and velocity divergence for GCG
\begin{eqnarray}
\dot{\delta}_{GCG}&=&-(1+w)(\theta_{GCG}+\frac{\dot{h}}{2})-3\mathcal{H}(c^{2}_{s}-w)\delta_{GCG}\\
\dot{\theta}_{GCG}&=&-\mathcal{H}(1-3c^{2}_{s})\theta_{GCG}+\frac{c^{2}_{s}}{1+w}k^{2}\delta_{GCG}-k^{2}\sigma_{GCG}
\end{eqnarray}
following the notation of Ma and Bertschinger \cite{ref:MB}. For the perturbation theory in gauge ready formalism, please see \cite{ref:Hwang}. The shear perturbation $\sigma_{GCG}=0$ is assumed and the adiabatic initial conditions are adopted in our calculation.

In this kind of unified dark fluid matter model, the averaging problem is involved for the non-linear perturbations. This problem comes from the fact $\langle p\rangle\neq p(\langle\rho\rangle)$ as pointed out by authors \cite{ref:stability}. Actually one can check that
\begin{equation}
\langle p\rangle=-\langle A/\rho^\alpha\rangle\neq -A/\langle\rho\rangle^\alpha=p(\langle\rho\rangle),
\end{equation}
in the case of $\alpha\neq 0$. But for the linear case ($\delta=\delta\rho/\langle \rho\rangle\ll 1$), one has \cite{ref:stability}
\begin{equation}
\langle p\rangle=-\langle A/\rho^\alpha\rangle=-A\langle \rho\rangle^{-\alpha}\langle (1-\alpha\delta)\rangle=-A\langle\rho\rangle^{-\alpha}.
\end{equation}
So we must be careful when we consider the large structure formation in this kind of unified dark fluid model. We are not going to discuss this problem deeply, because the main task of this paper is to discuss its effects on CMB and to constrain the model parameter space by using SN Ia, BAO and CMB data points. 

\section{Constraint method and results}\label{sec:method}

\subsection{Method and data points}

To constrain the parameter space, we use Markov Chain Monte Carlo (MCMC) method which is contained in a publicly available cosmoMC package \cite{ref:MCMC}, including the CAMB \cite{ref:CAMB} code to calculate the theoretical CMB power spectrum. We modified the code for the GCG as a unified fluid model with its perturbations included. The following $7$-dimensional parameter space  is adopted
\begin{equation}
P\equiv\{\omega_{b},\Theta_{S},\tau, \alpha, B_{s},n_{s},\log[10^{10}A_{s}]\}
\end{equation}
where $\omega_{b}=\Omega_{b}h^{2}$ is the physical baryon density, $\Theta_{S}$ (multiplied by $100$) is the ration of the sound horizon and angular diameter distance, $\tau$ is the optical depth, $\alpha$ and $B_{s}$ are two newly added model parameters related to GCG, $n_{s}$ is scalar spectral index, $A_{s}$ is the amplitude of of the initial power spectrum. Please notice that the current dimensionless energy density of GCG $\Omega_{GCG}$ is a derived parameter in a spatially flat ($k=0$) FRW universe. So, it is not included in the model parameter space $P$. The pivot scale of the initial scalar power spectrum $k_{s0}=0.05\text{Mpc}^{-1}$ is used. We take the following priors to model parameters: $\omega_{b}\in[0.005,0.1]$, $\Theta_{S}\in[0.5,10]$, $\tau\in[0.01,0.8]$, $\alpha\in[0,0.1]$, $B_{s}\in[0,1]$, $n_{s}\in[0.5,1.5]$ and $\log[10^{10}A_{s}]\in[2.7, 4]$. In addition, the hard coded prior on the comic age $10\text{Gyr}<t_{0}<\text{20Gyr}$ is imposed. Also, the weak Gaussian prior on the physical baryon density $\omega_{b}=0.022\pm0.002$ \cite{ref:bbn} from big bang nucleosynthesis and new Hubble constant $H_{0}=74.2\pm3.6\text{kms}^{-1}\text{Mpc}^{-1}$ \cite{ref:hubble} are adopted.

The total likelihood $\mathcal{L} \propto e^{-\chi^{2}/2}$ is calculated to get the distribution, here $\chi^{2}$ is given as
\begin{equation}
\chi^{2}=\chi^{2}_{CMB}+\chi^{2}_{BAO}+\chi^{2}_{SN}.
\end{equation}
The CMB data include temperature and polarization power spectrum from WMAP $7$-year data \cite{ref:lambda} as dynamic constraint. The geometric constraint comes from standard ruler BAO and standard candle SN Ia. For BAO, the values $\{r_{s}(z_{d})/D_{V}(0.2),r_{s}(z_{d})/D_{V}(0.5)\}$ and their inverse covariant matrix \cite{ref:BAO} are used. To use the BAO information, one needs to know the sound horizon at the redshift of drag epoch $z_{d}$. Usually, $z_{d}$ is obtained by using the accurate fitting formula \cite{ref:EH} which is valid if the matter scalings $\rho_{b}\propto a^{-3}$ and $\rho_{c}\propto a^{-3}$ are respected. Obviously, it is not true in our case. So, we find $z_{d}$ numerically from the following integration
\cite{ref:Hamann}
\begin{eqnarray}
\tau(\eta_d)&\equiv& \int_{\eta}^{\eta_0}d\eta'\dot{\tau}_d\nonumber\\
&=&\int_0^{z_d}dz\frac{d\eta}{da}\frac{x_e(z)\sigma_T}{R}=1
\end{eqnarray}
where $R=3\rho_{b}/4\rho_{\gamma}$, $\sigma_T$ is the Thomson cross-section and $x_e(z)$ is the fraction of free electrons. Then the sound horizon is
\begin{equation}
r_{s}(z_{d})=\int_{0}^{\eta(z_{d})}d\eta c_{s}(1+z).
\end{equation}
where $c_s=1/\sqrt{3(1+R)}$ is the sound speed. We use the substitution \cite{ref:Hamann}
\begin{equation}
d_z\rightarrow d_z\frac{\hat{r}_s(\tilde{z}_d)}{\hat{r}_s(z_d)}r_s(z_d),
\end{equation}
to obtain unbiased parameter and error estimates, where $d_z=r_s(\tilde{z}_d)/D_V(z)$, $\hat{r}_s$ is evaluated for the fiducial cosmology of Ref. \cite{ref:BAO}, and $\tilde{z}_d$ is obtained by using the fitting formula \cite{ref:EH} for the fiducial cosmology. Here $D_V(z)=[(1+z)^2D^2_Acz/H(z)]^{1/3}$ is the 'volume distance' with the angular diameter distance $D_A$.
The $557$ Union2 data with systematic errors are also included \cite{ref:Union2}. For the detailed description of SN, please see Refs. \cite{ref:Xu}.

\subsection{Fitting Results and Discussion}

\begin{center}
\begin{figure}[h]
\includegraphics[width=9cm]{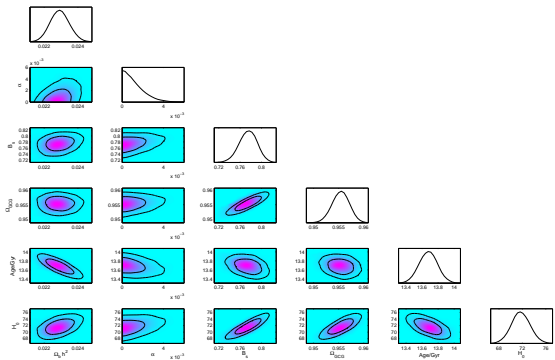}
\caption{The 1D marginalized distribution on individual parameter and 2D contours  with $68\%$ C.L. and $95\%$ C.L. by using CMB+BAO+SN data points.}\label{fig:contour}
\end{figure}
\end{center}

We generate $8$ independent chains in parallel and stop sampling by checking the worst e-values [the
variance(mean)/mean(variance) of 1/2 chains] $R-1$ of the order $0.01$. The calculated results of the model parameters and derived parameters are shown in Table. \ref{tab:results}, where the mean values with $1\sigma$ and $2\sigma$ regions from the combination WMAP+BAO+SN are listed. The minimum $\chi^2$ is $8010.420$ which is larger than that $\chi^2_{min}=8009.116$ for $\Lambda$CDM model with the same data sets combination. Correspondingly, the contour plots are shown in Figure \ref{fig:contour}.
\begingroup
\begin{table}
\begin{center}
\begin{tabular}{cc}
\hline\hline Prameters&Mean with errors\\ \hline
$\Omega_b h^2$ & $    0.0229_{-    0.000545-    0.00104}^{+    0.000548+    0.00110}$ \\
$\theta$ & $    1.0488_{-    0.00248-    0.00496}^{+    0.00248+    0.00483}$ \\
$\tau$ & $    0.0889_{-    0.00713-    0.0241}^{+    0.00656+    0.0249}$ \\
$\alpha$ & $    0.00126_{-    0.00126-    0.00126}^{+    0.000970+    0.00268}$ \\
$B_s$ & $    0.775_{-    0.0161-    0.0338}^{+    0.0161+    0.0307}$ \\
$n_s$ & $    0.980_{-    0.0145-    0.0274}^{+    0.0144+    0.0311}$ \\
$\log[10^{10} A_s]$ & $    3.0829_{-    0.0341-    0.0678}^{+    0.0337+    0.0679}$ \\
$\Omega_{GCG}$ & $    0.955_{-    0.00164-    0.00336}^{+    0.00166+    0.00319}$ \\
$Age/Gyr$ & $   13.677_{-    0.112-    0.218}^{+    0.110+    0.218}$ \\
$\Omega_b$ & $    0.04461_{-    0.00166-    0.00319}^{+    0.00164+    0.00336}$ \\
$z_{re}$ & $   10.498_{-    1.207-    2.454}^{+    1.200+    2.321}$ \\
$H_0$ & $   71.722_{-    1.525-    3.0408}^{+    1.535+    3.112}$ \\
\hline\hline
\end{tabular}
\caption{The mean values of model parameters with $1\sigma$ and $2\sigma$ errors from the combination WMAP+BAO+SN.}\label{tab:results}
\end{center}
\end{table}
\endgroup

From the Table \ref{tab:results} and Figure \ref{fig:contour}, one can clearly see that a tight constraint is obtained when the full information of CMB data is included. For the small values of $\alpha$, one finds that GCG model is very close to $\Lambda$CDM model.

\begin{center}
\begin{figure}[tbh]
\includegraphics[width=8.5cm]{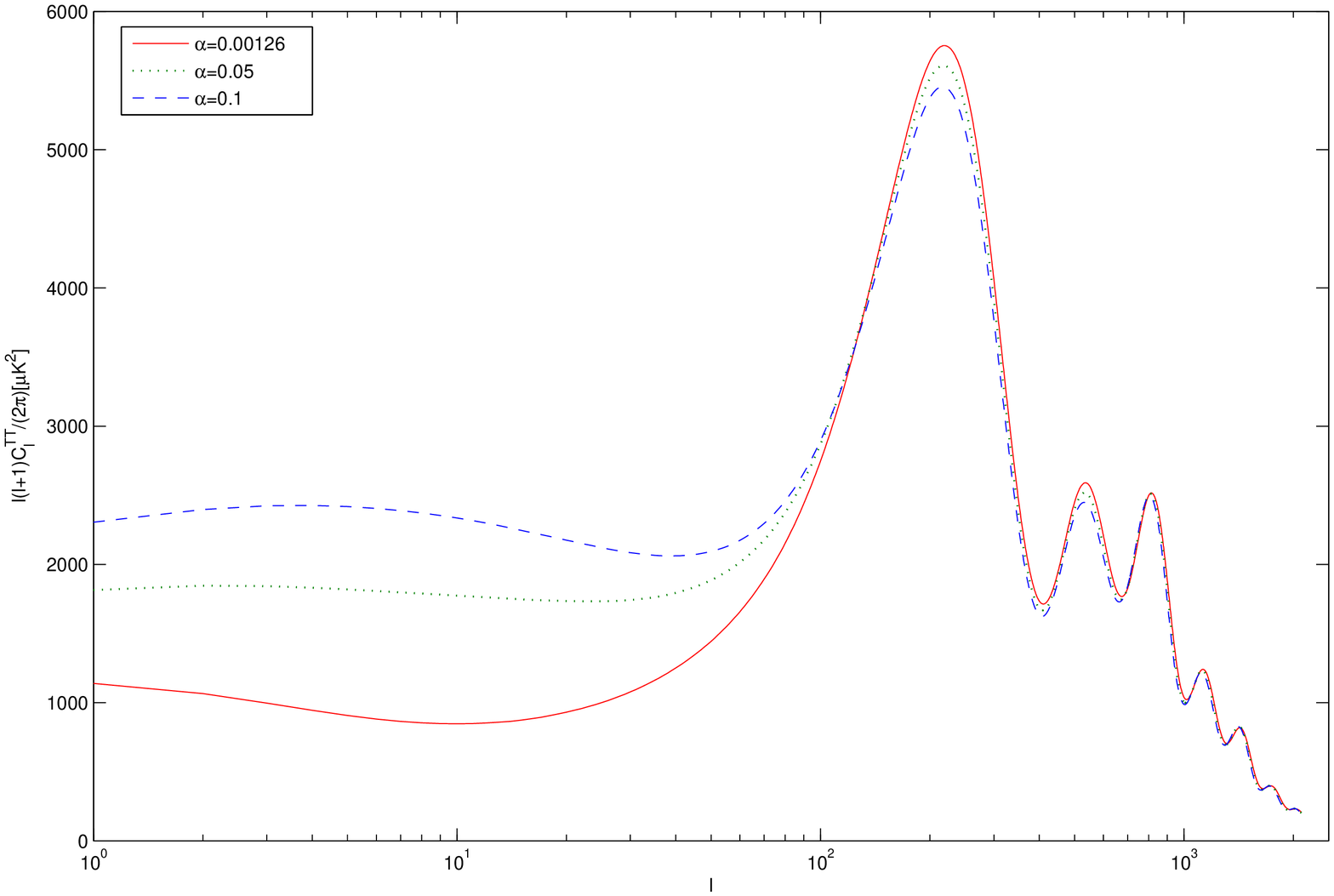}
\includegraphics[width=8.5cm]{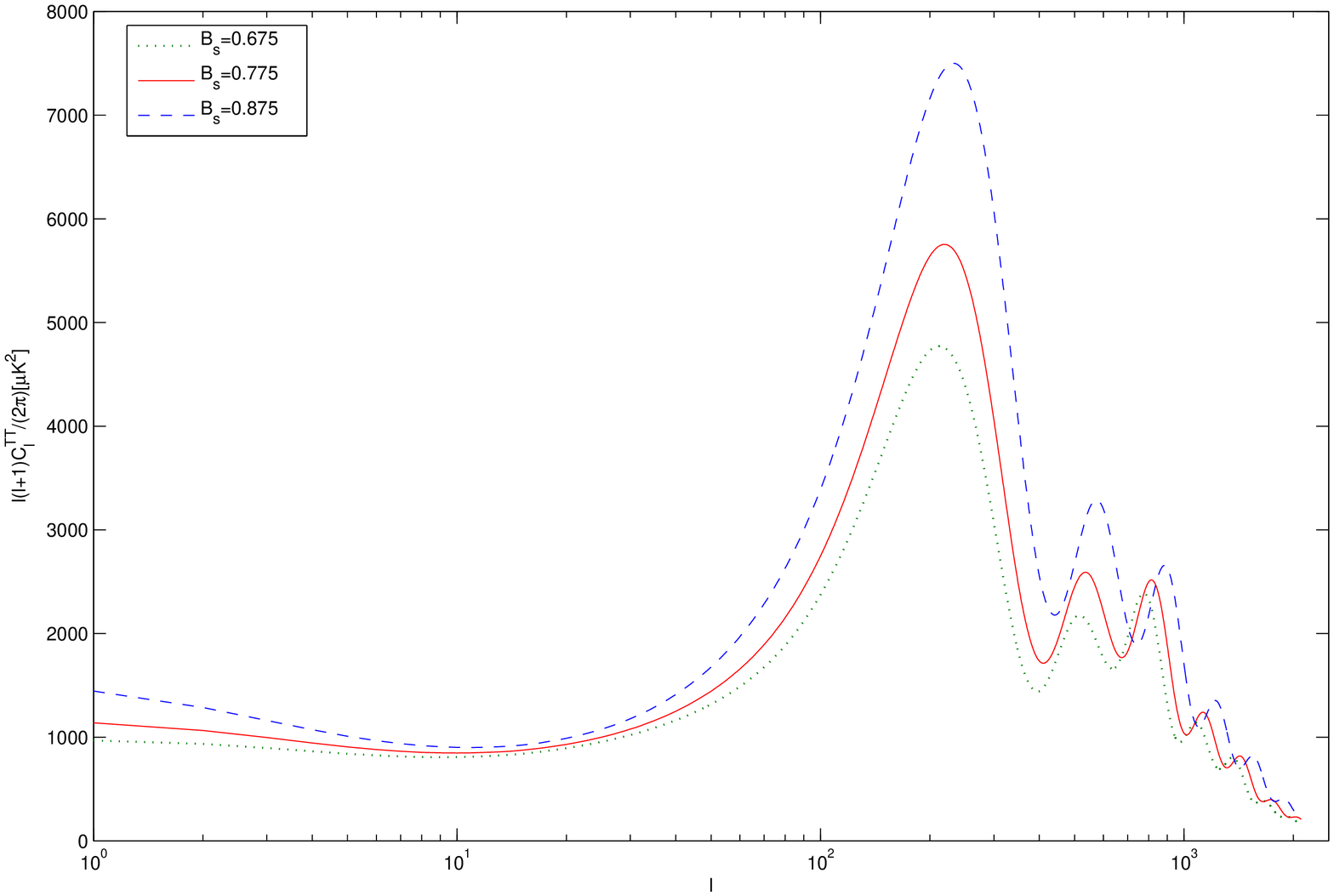}
\caption{The CMB $C^{TT}_l$ power spectrum v.s. multiple moment $l$. The panels show the effects of model parameters $\alpha$ and $B_s$ to CMB temperature anisotropic power spectra respectively, in each case the other relevant model parameters are fixed to the mean values as listed in Table \ref{tab:results}. }\label{fig:cls}
\end{figure}
\end{center}

\begin{center}
\begin{figure}[htb]
\includegraphics[width=9cm]{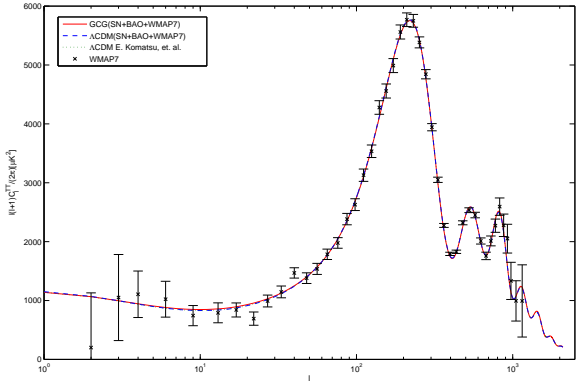}
\caption{The CMB $C^{TT}_l$ power spectrum v.s. multiple moment $l$, where the black dots with error
bars denote the observed data with their corresponding uncertainties from WMAP $7$-year results, the red solid line is for the unified dark fluid model with mean values as shown in Table \ref{tab:results}, the blue dashed line is for $\Lambda$CDM model with mean values for the same data points combination. And the green doted line is for $\Lambda$CDM model with mean values taken from \cite{ref:wmap7} with WMAP+BAO+$H_0$ constraint results.}\label{fig:mean}
\end{figure}
\end{center}

To understand the effects of model parameters to the CMB anisotropic power spectra, we plot the Figure \ref{fig:cls}, where one of two model parameters $\alpha$ and $B_s$ varies, where the other relevant parameters are fixed to their mean values as listed in Table \ref{tab:results}. The panels of Figure \ref{fig:cls} show the effect of parameter $\alpha$ and $B$ to CMB power spectra respectively. The model parameters $\alpha$ modifies the power law of the energy density of GCG, then it makes the gravity potential evolution at late epoch of the universe. As results, one can see Integrated SachsÐWolfe (ISW) effect on the large scale as shown in the left panel of Figure \ref{fig:cls}. In the early epoch, GCG behaves like cold dark matter with almost zero EoS and speed of sound $c^2_s$, therefore the variation of the values of $\alpha$ will change the ratio of energy densities of the effective cold dark matter and baryons. One can read the corresponding effects from the variation of the first and the second peaks of CMB power spectra. The parameter $B_s$ is related to the dimensionless density parameter of effective cold dark matter $\Omega_{c0}$. Decreasing the values of $B_s$, which is equivalent to increase the value of effective dimensionless energy density of cold dark matter, will make the equality of matter and radiation earlier, therefore the sound horizon is decreased. As a result, the first peak is depressed. The Figure \ref{fig:mean} shows CMB power spectra with mean values listed in Table \ref{tab:results} for GCG and $\Lambda$CDM model, where the black dots with error bars denote the observed data with their corresponding uncertainties from WMAP $7$-year results, the red solid line is for GCG with mean values as shown in Table \ref{tab:results}, the blue dashed line is for $\Lambda$CDM model with mean values for the same data points combination. And the green doted line is for $\Lambda$CDM model with mean values taken from \cite{ref:wmap7} with WMAP+BAO+$H_0$ constraint results. One can see that GCG can match observational data points and $\Lambda$CDM model well.

\begin{center}
\begin{figure}[tbh]
\includegraphics[width=8.5cm]{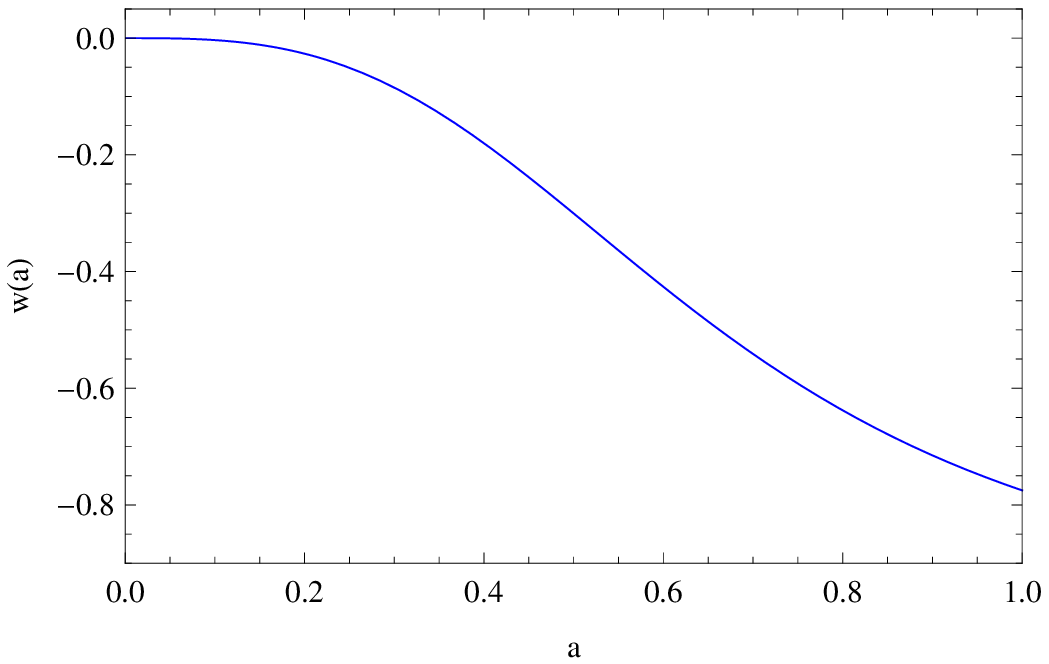}
\includegraphics[width=8.7cm]{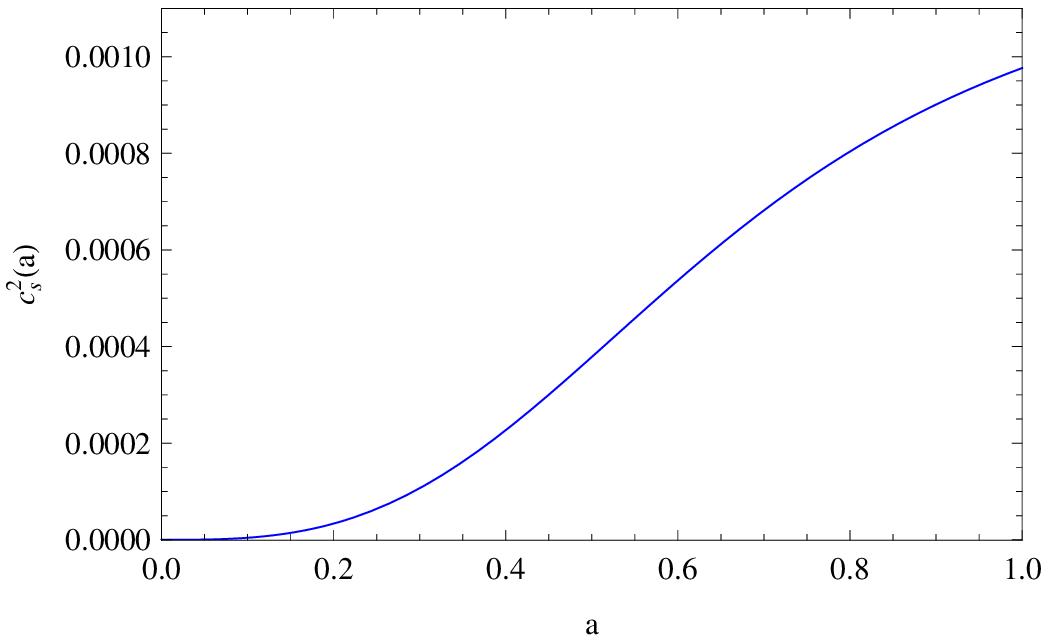}
\caption{The evolution of $w$ and $c^2_s$ with respect to scale factor $a$, where the model parameters are fixed to the mean values as listed in Table \ref{tab:results}. }\label{fig:csw}
\end{figure}
\end{center}

Now, we'd like to discuss the properties of GCG in the framework of unified dark fluid way, because the interpretation in the decomposition way is strongly decomposition dependent. The physical properties of GCG are described by its EoS $w$ and adiabatic sound speed $c^2_s$ which are determined by model parameters $B_s$ and $\alpha$. We can show the evolutions of $w$ and $c^2_s$ with respect to scale factor $a$ in Figure \ref{fig:csw}. From the left panel of Figure \ref{fig:csw}, one see that GCG behaves like cold dark matter with almost zero EoS and sound speed at early epoch ($a<0.2$). The almost zero sound speed is important for large scale structure formation. One can also read that it behaves like dark energy with EoS $w<0$ at late time, which pushes the universe into an accelerated phase. Here we do not consider the large structure formation in this unified GCG framework. In fact, it is not a easy issue. We leave it for the future work.

\section{Summary} \label{ref:conclusion}

In summary, we perform a global fitting on GCG model, which is treated as a unified dark matter and dark energy model, by using MCMC method with the combination of the full CMB, BAO and  SN Ia data points. As a contrast to the reports in the literatures, we take GCG as an entire energy component and without any decomposition. Tight constraint is obtained as shown in Table \ref{tab:results} and Figure \ref{fig:contour}. The GCG model can match observational data points and $\Lambda$CDM model well. For the small values of model parameter $\alpha$, one can conclude that MCG model is very close to $\Lambda$CDM model. And currently available data sets of CMB, BAO and SN Ia can not distinguish GCG model from $\Lambda$CDM model.  

\section{Acknowledgements}
We thank an anonymous referee for helpful improvement of this paper. L. Xu's work is supported by the Fundamental Research Funds for the Central Universities (DUT10LK31) and (DUT11LK39). J. Lu's
work is supported by  the National Natural Science Foundation of
China (11147150) and the Natural Science Foundation of Education
Department of Liaoning Province, China (L2011189).

\end{document}